\documentclass[twocolumn,amsmath,amssymb,aps,superscriptaddress]{revtex4-2}
\pdfoutput=1
\usepackage{package}

\begin{document}
\title{Scalable circuit depth reduction in feedback-based quantum optimization with a quadratic approximation}

\author{Don Arai}%
\thanks{These two authors contributed equally. Correspondence: u966637k@ecs.osaka-u.ac.jp (D.A.), okada.ken.qiqb@osaka-u.ac.jp (K.N.O.)}
\affiliation{%
  Graduate School of Engineering Science, Osaka University, 1-3 Machikaneyama, Toyonaka, Osaka 560-8531, Japan
}
\author{Ken N. Okada}%
\thanks{These two authors contributed equally. Correspondence: u966637k@ecs.osaka-u.ac.jp (D.A.), okada.ken.qiqb@osaka-u.ac.jp (K.N.O.)}
\affiliation{%
  Center for Quantum Information and Quantum Biology,
  Osaka University, 1-2 Machikaneyama, Toyonaka 560-0043, Japan
}%
\author{Yuichiro Nakano}%
\affiliation{%
  Graduate School of Engineering Science, Osaka University, 1-3 Machikaneyama, Toyonaka, Osaka 560-8531, Japan
}
\author{Kosuke Mitarai}%
\affiliation{%
  Graduate School of Engineering Science, Osaka University, 1-3 Machikaneyama, Toyonaka, Osaka 560-8531, Japan
}
\affiliation{%
  Center for Quantum Information and Quantum Biology,
  Osaka University, 1-2 Machikaneyama, Toyonaka 560-0043, Japan
}%
\author{Keisuke Fujii}%
\affiliation{%
  Graduate School of Engineering Science, Osaka University, 1-3 Machikaneyama, Toyonaka, Osaka 560-8531, Japan
}
\affiliation{%
  Center for Quantum Information and Quantum Biology,
  Osaka University, 1-2 Machikaneyama, Toyonaka 560-0043, Japan
}%
\affiliation{%
  RIKEN Center for Quantum Computing (RQC),
  Hirosawa 2-1, Wako, Saitama 351-0198, Japan
}%

\date{\today}

\begin{abstract}
Combinatorial optimization problems are one of the areas where near-term noisy quantum computers may have practical advantage against classical computers.
  Recently a novel feedback-based quantum optimization algorithm has been proposed by Magann \textit{et al} \cite{magann2022feedback}.
  The method explicitly determines quantum circuit parameters by feeding back measurement results thus avoids classical parameter optimization that is known to cause significant trouble in quantum approximate optimization algorithm, the well-studied near-term algorithm.
  Meanwhile, a significant drawback of the feedback-based quantum optimization is that it requires deep circuits, rendering the method unsuitable to noisy quantum devices.
  In this study we propose a new feedback law for parameter determination by introducing the second-order approximation with respect to time interval, a hyperparameter in the feedback-based quantum optimization.
  This allows one to take larger time interval, leading to acceleration of convergence to solutions.
  In numerical simulations on the maximum cut problem we demonstrate that our proposal significantly reduces circuit depth, with its linear scaling with the problem size smaller by more than an order of magnitude.
  We expect that the new feedback law proposed in this work may pave the way for feedback-based quantum optimization with near-term noisy quantum computers.
\end{abstract}

\maketitle

\section{Introduction}
Combinatorial optimization problems appear in a wide range of fields in basic science and industrial application, e.g., protein folding and traveling salesman problems.
It is known that solving combinatorial optimization problems is NP-hard in general, while various approximation algorithms have been developed in discrete mathematics \cite{korte2000combinatorial}.
More recently, emergence of quantum annealing machines and noisy intermediate-scale quantum (NISQ) \cite{preskill2018quantum, ichikawa2023comprehensive} computers has led to extensive studies to solve combinatorial optimization problems harnessing the effects of quantum superposition and interference effects \cite{das2008colloquium, blekos2024review, abbas2023quantum}.

Quantum approximate optimization algorithm (QAOA) \cite{farhi2014quantum} has attracted much attention as an algorithm for solving combinatorial optimization problems using NISQ devices.
QAOA is a hybrid quantum-classical algorithm \cite{peruzzo2014variational, mcclean2016theory}, where one prepares states on quantum circuits with variational parameters and optimizes the parameters on a classical computer through minimization of a cost function obtained by measurement \cite{cerezo2021variational,bharti2022noisy}.
While QAOA has been considered suitable for NISQ devices because of relatively shallow circuits, parameter optimization becomes a major obstacle because of sampling costs during the optimization as well as barren plateaus appearing in the cost function landscape \cite{mcclean2018barren,larocca2022diagnosing}.

Recently a distinct type of quantum optimization algorithm based on quantum control theories \cite{PhysRevLett.69.2172, sugawara2003general, grivopoulos2003lyapunov} has been proposed by Magann \textit{et al} \cite{magann2022feedback, magann2022lyapunov}.
The algorithm, named Feedback-based ALgorithm for Quantum OptimizatioN (FALQON), iteratively adds quantum circuit layers with parameters determined by measurement results on the latest quantum state.
Most importantly, unlike QAOA, it does not require parameter optimization by a classical computer.
Magann \textit{et al.} have shown numerically that FALQON gives almost exact solutions to maximum cut (MAX-CUT) problems on small instances \cite{magann2022feedback, magann2022lyapunov}.
They have also found out that the number of layers needed almost linearly scales with graph size.
Later on, the same feedback approaches have been applied to ground state preparation in quantum many-body problems \cite{larsen2023feedback}.
Its convergence has been further improved with an additional control field inspired by counterdiabatic control protocol \cite{malla2024feedback}.

Despite the numerical success, FALQON still needs to be improved to be implemented on actual NISQ devices. 
The biggest issue is its large circuit depth.
For example, for unweighted MAX-CUT problems on 3-regular graphs, circuit depth of FALQON becomes almost two orders of magnitude larger than that of QAOA to achieve the same level of accuracy \cite{magann2022feedback, zhou2020quantum}.

With the aim of reducing circuit depth in FALQON towards implementation in the NISQ era, here we propose a modified feedback law for determining parameters by incorporating higher-order expansions.
More specifically, we improve the feedback law of the original FALQON \cite{magann2022lyapunov, magann2022feedback} which depends on the first-order Taylor expansion of the cost function difference per step to a second-order version.
While the first-order expansion provides a simple update law, it imposes a restriction that the action of each quantum circuit layer added at an iteration must be small enough, meaning that the quantum state can only evolve slowly to the solution.
The use of the second-order expansion relaxes this restriction to a certain extent and allows us to speed up the convergence.
We apply this method to the MAX-CUT problem on 3-regular graphs and numerically demonstrate that the coefficient of the linear scaling of the circuit depth with graph size is reduced by more than an order of magnitude as compared to FALQON using the conventional feedback law.
Our proposal of the second-order feedback law could be expected to open the way for implementation of feedback-based quantum algorithms on NISQ computers.

The rest of this paper is organized as follows.
In Sec. \ref{FALQON_review} we present a review of FALQON proposed by Magann \textit{et al.} \cite{magann2022feedback} and its limitations.
In Sec. \ref{so-falqon} we propose a new feedback law based on a second-order approximation that could mitigate the limitations explained in Sec. \ref{FALQON_review}.
In Sec. \ref{numerics} we show numerical simulations of FALQON with the new feedback law proposed in Sec. \ref{so-falqon} applied to the MAX-CUT problem.
Lastly, in Sec. \ref{conclusion}, we summarize our results and provide potential future studies.

\section{Feedback-based Algorithm for Quantum Optimization (FALQON)}\label{FALQON_review}

\subsection{Theory} \label{theory}
In this section, we provide an overview of FALQON \cite{magann2022feedback}.
As in the cases of quantum annealing and QAOA, the cost function of a given combinatorial optimization problem is encoded into the problem Hamiltonian $H_p$.
Inspired by the concept of quantum Lyapunov control \cite{PhysRevLett.69.2172, sugawara2003general, grivopoulos2003lyapunov}, FALQON attempts to minimize the expectation value of $H_p$ by realizing time evolution of the system considered in a desired manner through measurement-based feedback control.
Based on the works of Magann \textit{et al.} \cite{magann2022feedback, magann2022lyapunov}, we first review FALQON in the continuous time framework and then explain how it can be implemented on quantum circuits.

FALQON uses dynamics under a time-dependent Hamiltonian in the form of
\begin{align}
  i\frac{d}{dt}\ket{\psi(t)} = (H_p + H_d\beta(t))\ket{\psi(t)}, \label{dynamics}
\end{align}
where $H_d$ is a driver Hamiltonian and $\beta(t)\in \mathbb{R}$ is a time-dependent real coefficient that is to be determined sequentially.
To minimize the expectation value of $H_p$, $E_p(t)=\ev{H_p}{\psi(t)}$, we consider designing $\beta(t)$ so that $E_p(t)$ always decreases with time, i.e., $\frac{d}{dt}E_p(t) \leq 0,~ \forall t \geq 0.$
Eq.~\eqref{dynamics} yields
\begin{align}
  \frac{d}{dt}E_p(t) = A(t)\beta(t), \label{betaA}
\end{align}
where $A(t)$ is defined as
\begin{align}
  A(t) = \bra{\psi(t)}i[H_d,H_p]\ket{\psi(t)}. \label{A}
\end{align}
Therefore, if one determines $\beta(t)$ so that the right-hand side of Eq.~\eqref{betaA} is less than 0, one can drive time evolution that allows $E_p(t)$ to monotonically decrease.
While there is much flexibility in determination of such $\beta(t)$, one of the simplest choices is $\beta(t)$ = $-A(t)$ as employed in refs. \cite{magann2022feedback, magann2022lyapunov}.
In practice, $\beta(t)$ is determined by measurement on the state small time earlier as $\beta(t)$ = $-A(t-\tau)$.

\begin{figure}[t]
  \centering
  \includegraphics[scale = 0.37]{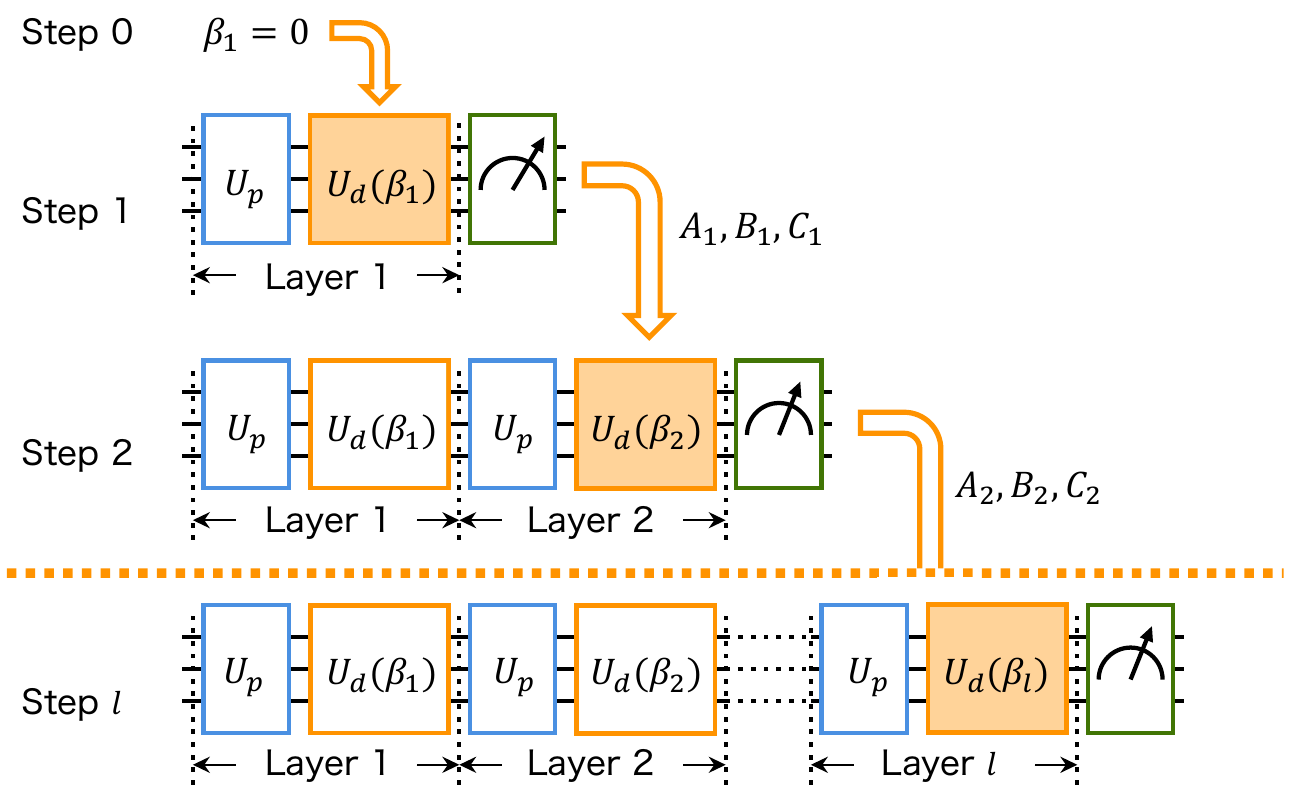}
  \caption{Schematic diagram of the iterative procedures in FALQON. In step $k$ a new circuit parameter $\beta_k$ is added and determined through feedback laws using expectation values $A_{k-1}$, $B_{k-1}$, and $C_{k-1}$ obtained by measurement in step $k-1$. In FALQON proposed by Magann \textit{et al.} \cite{magann2022feedback} (FO-FALQON) the feedback law is simply expressed by Eq. \eqref{betak} only involving $A_{k-1}$. In FALQON proposed in this work (SO-FALQON) the feedback law is expressed by Eq. \eqref{newbetak} additionally involving $B_{k-1}$ and $C_{k-1}$.}
  \label{outline}
\end{figure}

To implement FALQON on the quantum circuit model, the continuous time dynamics described above is discretized via the Trotter decomposition.
We define $U_d(\beta_k) = e^{-i\beta_kH_d\Delta t}$ and $U_p = e^{-iH_p\Delta t}$ with small time interval $\Delta t$ and control parameter $\beta_k$.
The dynamics of Eq. \eqref{dynamics} can be approximated by the alternate application of $U_d(\beta_k)$ and $U_p$ as $U_{k} = U_d(\beta_k)U_p\dots U_d(\beta_1)U_p$.
FALQON iteratively determines $\beta_k$ and thereby drives the state to a solution as follows [Fig.~\ref{outline}].
First, we prepare an initial state of $\ket{\psi_0}$ and set $\beta_1=0$ (step 0).
In step 1, we generate the state $\ket{\psi_1}$ as $\ket{\psi_1} = U_d(\beta_1)U_p\ket{\psi_0}=U_p\ket{\psi_0}$
and obtain $A_1 = \ev{i[H_d,H_p]}{\psi_1}$.
Then we determine $\beta_2$ via the feedback law as $\beta_2=-A_1$.
In step $k$, one generates the state $\ket{\psi_k}$ by applying $U_d(\beta_k)U_p$ to the previous state 
$\ket{\psi_{k-1}}$, where $\beta_{k}$ is determined by the feedback law 
\begin{align}
   \beta_{k}=-A_{k-1} \label{betak}
\end{align}
with
\begin{align}
   A_k = \ev{i[H_d,H_p]}{\psi_k} \label{Ak}.
\end{align}
As long as $\Delta t$ is sufficiently small, the expectation value of $H_p$ monotonically decreases as $\ev{H_p}{\psi_1} \geq \ev{H_p}{\psi_2} \geq \dots \geq \ev{H_p}{\psi_k}$.
FALQON minimizes the energy expectation value by increasing the number of steps (layers) to obtain the solution to a given combinatorial optimization problem.

\subsection{Breakdown of optimization for large $\Delta t$} \label{demo}

\begin{figure}[t]
  \centering
  \includegraphics[scale = 0.42]{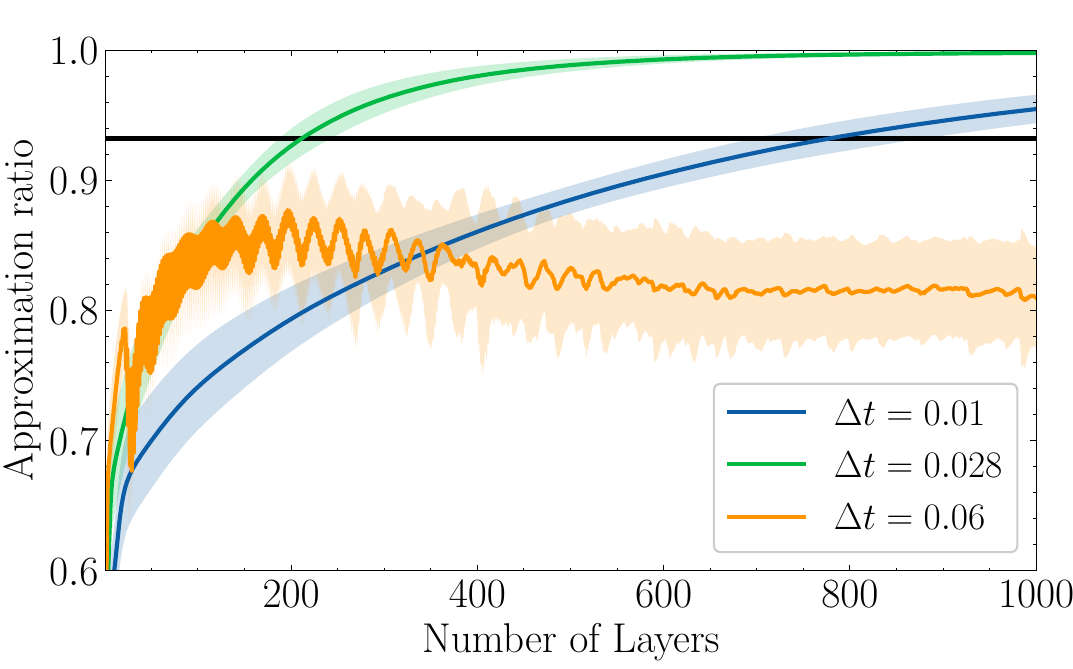}
  \caption{Numerical simulation of FALQON using the feedback law in Eq. \eqref{betak}, proposed by Magann \textit{et al} \cite{magann2022feedback}. The approximation ratio is shown as a function of number of layers (steps) for $n=12$ with different $\Delta t$. The lines and shaded areas represent the means and standard deviations over different graphs, respectively. The black line represents the reference value $r_{GW} = 0.932$ guaranteed by the Goemans-Williamson algorithm \cite{goemans1995improved}, the best classical algorithm with a performance guarantee.}
  \label{reproduction}
\end{figure}

Here, we demonstrate the breakdown of the original feedback law, Eq. \eqref{betak}, when using large $\Delta t$.
To this end, we use the MAX-CUT problem of 3-regular graphs following the original work \cite{magann2022feedback}.
The MAX-CUT problem asks us to find a ground state of 
\begin{align}
   H_p = -\frac{1}{2}\sum_{(i, j)\in E} (1-Z_iZ_j),
\end{align}
for a given graph $G=(V,E)$ where $V$ and $E$ are sets of vertices and edges.
We adopt the driver Hamiltonian $H_d = -\sum_{i=1}^{n} X_i$ and accordingly set the initial state as $\ket{\psi_0}=\ket{+}^{\otimes n}$, where $n$ is the number of vertices.

Figure~\ref{reproduction} shows the evolution of the approximation ratio defined as $r=\ev{H_p}/E_{\min}$, where $E_{\min}$ is the ground state energy of $H_p$, during FALQON iteration at $\Delta t$ = 0.01, 0.028, and 0.06 for $n = 12$.
The approximation ratio increases much faster when $\Delta t$ is increased from 0.01 to 0.028.
However, when $\Delta t$ is increased up to 0.06, the approximation ratio does not increase monotonically.
This is because $\Delta t$ is not small enough to retain monotonic decrease in the energy originally guaranteed in the continuous time framework.
We aim to resolve this breakdown using a higher-order feedback law in the next section.

\section{Second-order FALQON}\label{so-falqon}
Here we propose a new feedback law for determining control parameters $\beta_k$ that enables faster convergence of FALQON.
In the iterative procedure of FALQON [Fig.~\ref{outline}], the difference in the expectation value of $H_p$ between steps $k$ and $k-1$ is expressed as
\begin{eqnarray}
  \Delta \ev{H_p}_k &=& \ev{H_p}_{k} - \ev{H_p}_{k-1} \nonumber\\
  &=& \ev*{U_p^{\dagger}U_d^{\dagger}(\beta_k)H_pU_d(\beta_k)U_p}{\psi_{k-1}} \nonumber\\
  &\ &- \ev{H_p}{\psi_{k-1}}.\label{eq1}
\end{eqnarray}
By applying the Taylor expansion to $U_d(\beta_k)$ and $U_p$ up to the second order with respect to $\Delta t$, one obtains $U_d(\beta_k) \simeq 1 - i\beta_kH_d\Delta t - \frac{1}{2}(\beta_kH_d\Delta t)^2$ and $U_p \simeq 1  - i\ H_p\Delta t - \frac{1}{2}(H_p\Delta t)^2$.
Then one can approximate Eq.~\eqref{eq1} as
\begin{align}
  \Delta \ev{H_p}_k &\simeq \Delta t \beta_k \ev{i\qty[H_d, H_p]}{\psi_{k-1}} \nonumber\\
  &+ \qty(\Delta t)^2\qty(\beta_k)^2\ev*{\frac{1}{2}\qty[\qty[H_d,H_p], H_d]}{\psi_{k-1}}\nonumber\\
  &+ \qty(\Delta t)^2\beta_k\ev*{[\qty[H_d,H_p],H_p]}{\psi_{k-1}}\nonumber\\
  &+ \order{(\Delta t)^3}. \label{eq2}
\end{align}
The first-order term with respect to $\Delta t$ in Eq.~\eqref{eq2} is equivalent to the time derivative of the energy discussed in the continuous limit in Sec.~\ref{theory}, Eq.~\eqref{betaA}.
This means that in FALQON $\beta_k$ is determined so that $\Delta \ev{H_p}_k$ decreases up to the first order with respect to $\Delta t$ \cite{magann2022lyapunov}.

In this work we propose a new feedback law that determines $\beta_k$ so that $\ev{H_p}_k$ decreases up to the second order with respect to $\Delta t$. 
One can expect that incorporating the second-order terms to the approximation of $\Delta \ev{H_p}_k$ guarantees monotonic decrease in $\ev{H_p}_k$ for larger $\Delta t$.
In the following, we call FALQON using the new feedback law second-order FALQON (SO-FALQON) and FALQON using the feedback law in Eq. \eqref{betak} first-order FALQON (FO-FALQON).
Since Eq.~\eqref{eq2} can be expressed as a quadratic function with respect to $\beta_k$ as $\Delta \ev{H_p}_k \simeq (\Delta t)^2 B_{k-1} \qty(\beta_k+\frac{A_{k-1}+ \Delta tC_{k-1}}{2\Delta t B_{k-1}})^2 - \frac{1}{4B_{k-1}}\qty(A_{k-1}+ \Delta tC_{k-1})^2$, we define the feedback law of SO-FALQON as
\begin{align}
  \beta_k = -\frac{A_{k-1}+ \Delta tC_{k-1}}{2\Delta t B_{k-1}} \label{newbetak}.
\end{align}
Here we define new expectation values $B_k$ and $C_k$ as
\begin{align}
    B_k = \ev*{\frac{1}{2}\qty[\qty[H_d,H_p], H_d]}{\psi_{k-1}},\\
    C_k = \ev*{[\qty[H_d,H_p],H_p]}{\psi_{k-1}}.
\end{align}
For $B_k>0$ Eq.~\eqref{newbetak} always gives a negative $\Delta \ev{H_p}_k$ and corresponds to the minimum of $\Delta \ev{H_p}_k$.
Meanwhile, for $B_k<0$ one can guarantee a negative $\Delta \ev{H_p}_k$ by changing the sign of the right-hand side of Eq.~\eqref{newbetak}.

When implementing SO-FALQON, one may need to be careful that the absolute value of Eq. \eqref{newbetak} does not become too large.
This is because $|\beta_k|\norm{H_d}\Delta t\ll 1$ has to be satisfied for the approximation in Eq. \eqref{eq2}, and Eq. \eqref{newbetak} could more easily violate the condition than Eq. \eqref{betak} due to the denominator including $\Delta t$.
Therefore one might be encouraged to put some cap on $\beta_k$ in SO-FALQON as done in the following simulations.

\section{numerical simulations}\label{numerics}
\subsection{Method}\label{method}
In this section we compare the performance of SO-FALQON and FO-FALQON by numerical simulation.
As in the previous study \cite{magann2022feedback}, we consider the MAX-CUT problem on unweighted 3-regular graphs.
The number of vertices is $n \in \qty{8, 10, 12, 14, 16}$.
We study all nonisomorphic graphs for $n \in \qty{8, 10}$ and 50 randomly chosen nonisomorphic graphs for $n \in \qty{12, 14, 16}$.
We evaluate means and standard deviations over the graphs for each $n$.
We refer to Sec. \ref{demo} for other settings such as $H_p$, $H_d$, and $\ket{\psi_0}$.
Numerical simulations were performed using a fast quantum circuit simulator Qulacs \cite{suzuki2021qulacs}.

To cap the value of $\beta_k$ in SO-FALQON as noted in Sec. \ref{so-falqon}, we make hybrid usage of the second-order and first-order feedback laws.
Concretely, we adopt Eq. \eqref{newbetak} in cases that the absolute value of Eq. \eqref{newbetak} is smaller than that of Eq. \eqref{betak} and otherwise use Eq. \eqref{betak}.
Note that we have found that $B_k>0$ always holds in our simulations.
We refer to Appendix for simulations of SO-FALQON without introducing the cap by Eq. \eqref{betak}.

\subsection{Results}
\begin{figure}[t]
  \centering
  \includegraphics[scale = 0.4]{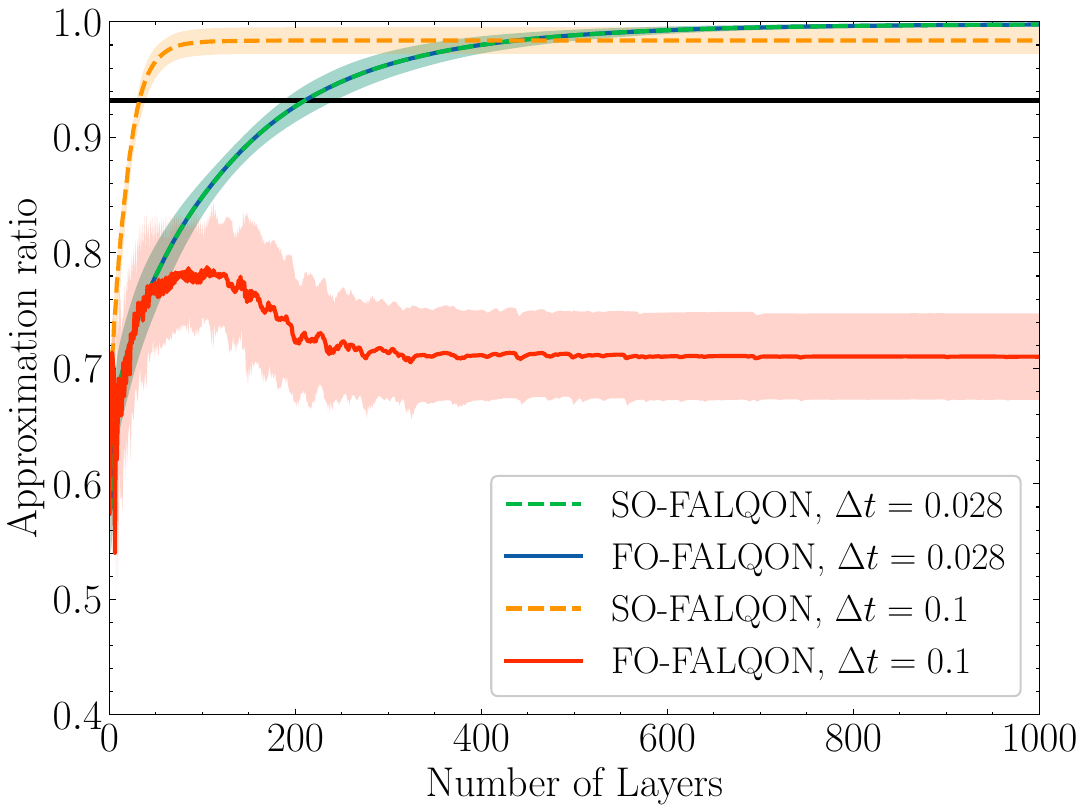}
  \caption{Numerical simulation of SO-FALQON along with FO-FALQON with different $\Delta t$ for $n = 12$.}
  \label{second}
\end{figure}

In Fig.~\ref{second} we show the results of SO-FALQON along with FO-FALQON with $\Delta t = 0.028, 0.1$ for $n = 12$.
$\Delta t = 0.028$ corresponds to the critical $\Delta t$ below which the approximation ratio of FO-FALQON monotonically increases, denoted as $\Delta t_c^\qty(1)$ hereafter.
For $\Delta t = 0.028$, the approximation ratio of both FALQONs monotonically increases and converges to 1.
SO-FALQON exhibits the same curve as FO-FALQON because the absolute value of Eq. (\refeq{betak}) is always smaller than that of Eq. (\ref{newbetak}) and hence selected every step.
For $\Delta t = 0.1$, the approximation ratio of FO-FALQON fails to increase monotonically and stays around a low value of 0.71 for large numbers of layers $L$.
On the other hand, the approximation ratio of SO-FALQON still increases monotonically and approaches as high as 0.98.
The robust success of SO-FALQON for larger $\Delta t$ can be attributed to the new feedback law in Eq. \eqref{newbetak}, which allows monotonic decrease in $\ev{H_p}_k$ for larger $\Delta t$ because of the second-order approximation of $\Delta \ev{H_p}_k$ with respect to $\Delta t$.
In Appendix we have confirmed that without introducing the cutoff by Eq. (\refeq{betak}) SO-FALQON still yields monotonic increase in the approximation ratio for $\Delta t=0.1$ in most cases.

\begin{figure}[t]
  \centering
  \includegraphics[scale = 0.4]{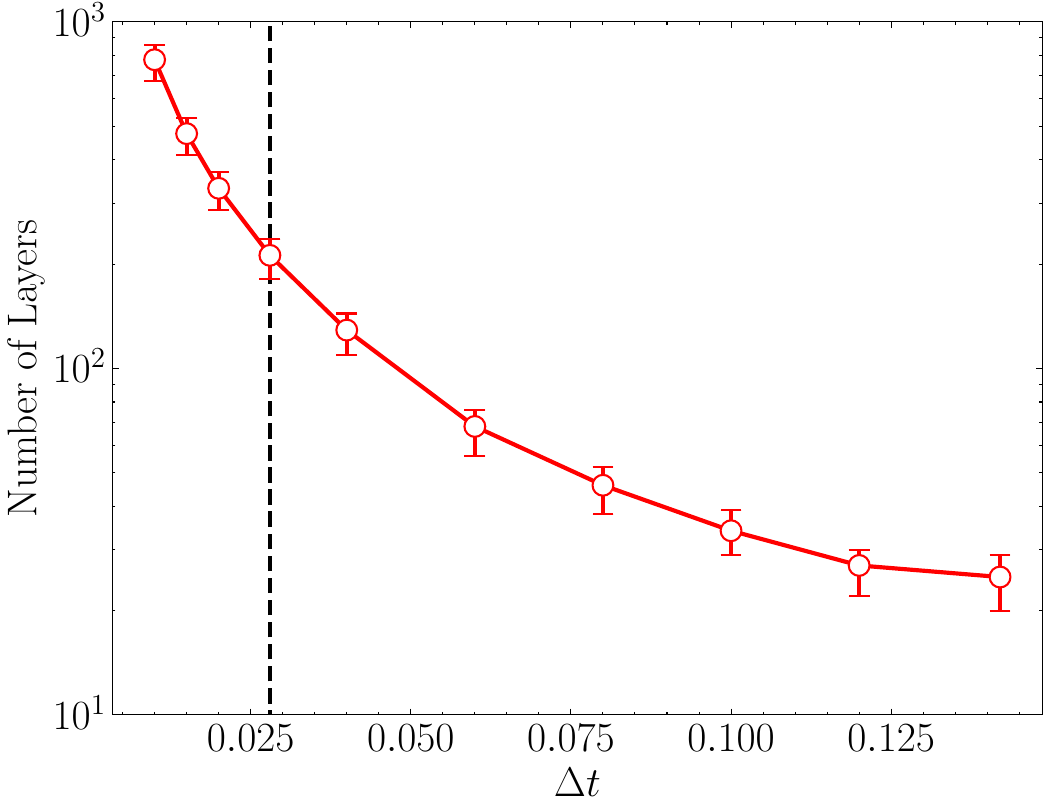}
  \caption{$\Delta t$ dependence of the number of layers required for SO-FALQON to reach $r_{GW} = 0.932$ for $n = 12$. The dashed line represents $\Delta t^{(1)}_c$. The region in $\Delta t_c \leq \Delta t^{(1)}_c$ can be covered by FO-FALQON as well, whereas $\Delta t > \Delta t^{(1)}_c$ can be only reached by SO-FALQON.}
  \label{dt_L}
\end{figure}

Figure~\ref{dt_L} shows the number of layers required for SO-FALQON to reach the approximation ratio guaranteed by the classical Goemans-Williamson algorithm \cite{goemans1995improved} $r_{GW} = 0.932$ as $\Delta t$ is varied.
The dashed line represents $\Delta t^{\qty(1)}_c$.
The largest $\Delta t$ among the data points in Fig.~\ref{dt_L} corresponds to $\Delta t^{\qty(2)}_c$, the critical value of $\Delta t$ below which SO-FALQON achieves monotonic increase in approximation ratio.
Therefore the region in $\Delta t \leq \Delta t^{\qty(1)}_c$ can also be covered by FO-FALQON, whereas the region in $\Delta t^{\qty(1)}_c < \Delta t \leq \Delta t^{\qty(2)}_c$ can be only reached by SO-FALQON.
Figure~\ref{dt_L} indicates that the number of layers required drastically decreases with $\Delta t$.
Comparing the data for $\Delta t=\Delta t^{\qty(1)}_c$ and $\Delta t=\Delta t^{\qty(2)}_c$, one can see that in SO-FALQON the number of layers required is reduced by nearly an order of magnitude as compared to FO-FALQON.

\subsection{Scaling comparison of SO-FALQON and FO-FALQON}
\begin{figure}[t]
  \centering
  \includegraphics[scale = 0.4]{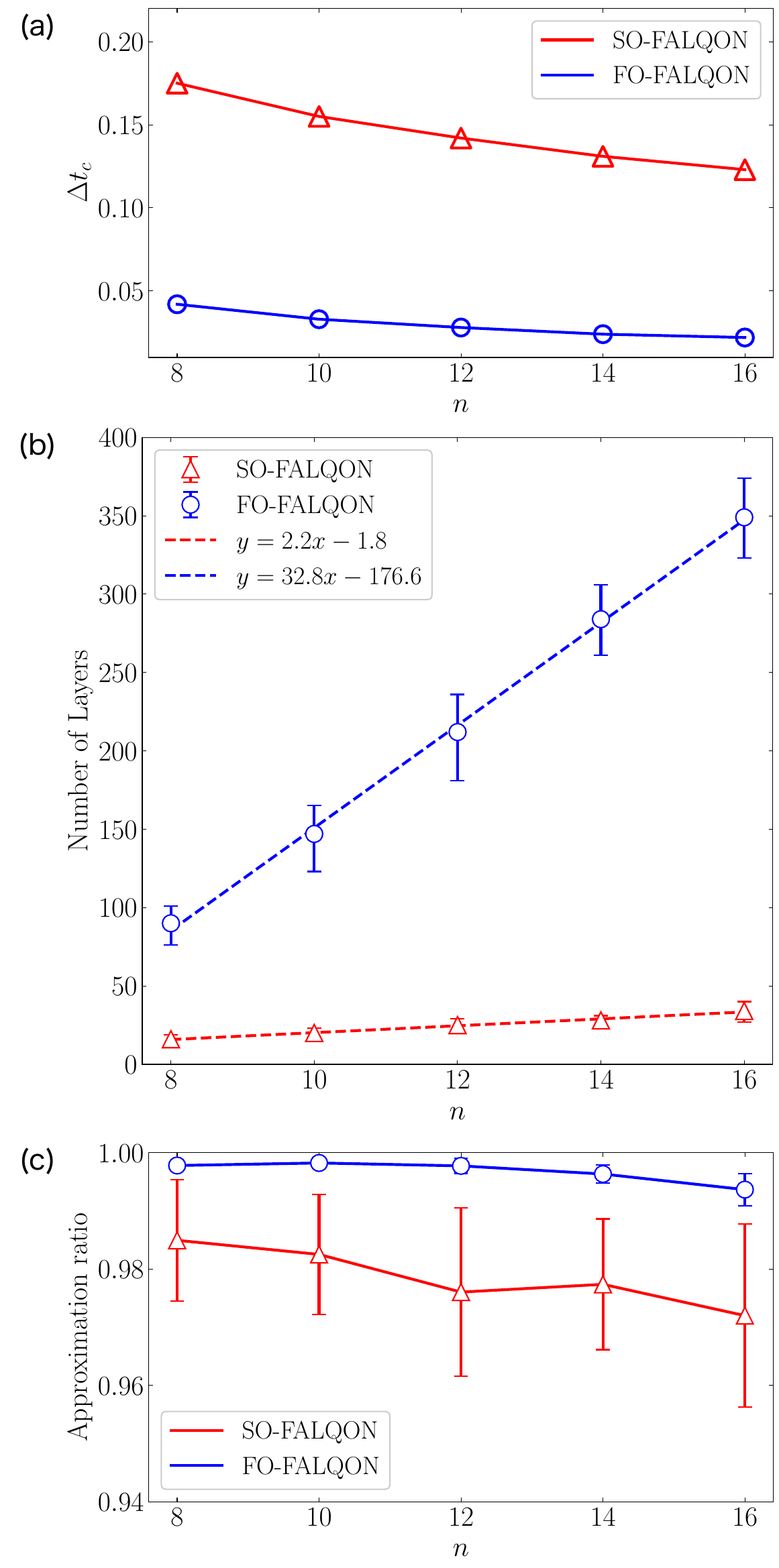}
  \caption{Graph size dependence of SO-FALQON and FO-FALQON. (a) Critical value of $\Delta t$, $\Delta t_c$. (b) Number of layers required to obtain the classically guaranteed approximation ratio when $\Delta t=\Delta t_c$. (c) Approximation ratio at $L=1000$.}
  \label{compare}
\end{figure}
Finally, we compare the performance of SO-FALQON and FO-FALQON through the graph size dependence.
In Figs.~\ref{compare}(a) and \ref{compare}(b), we present the critical value of $\Delta t$ that allows monotonic increase in the approximation ratio, $\Delta t_c$, and the number of layers required at $\Delta t=\Delta t_c$ to reach the classically guaranteed approximation ratio, respectively.
Figure~\ref{compare}(a) shows that although $\Delta t_c$ decreases with $n$ for both FALQONs, it is significantly larger for SO-FALQON than for FO-FALQON.
Figure~\ref{compare}(b) indicates that while the number of layers required almost linearly increases with $n$ it is drastically smaller in SO-FALQON than in FO-FALQON. 
The reduction in the number of layers originates from increase in $\Delta t_c$, allowed by the new feedback law incorporating the second-order terms in approximating the energy difference. 
As we fit each data with a linear function $y=ax+b$, we obtain $a=2.2$ for SO-FALQON and $a=32.8$ for FO-FALQON as represented by dashed lines in Fig.~\ref{compare}(b).
This implicates that the number of layers required is scalably reduced by more than an order of magnitude.

Lastly we note some drawback to increase in $\Delta t_c$.
Figure~\ref{compare}(c) shows the approximation ratio in the large $L$ limit ($L=1000$) as a function of $n$.
While the approximation ratio tends to decrease with $n$ for both FALQONs, SO-FALQON overall shows $r\sim 0.98$, smaller than $r\sim 0.995$ for FO-FALQON.
The reason why the saturation value slightly decreases in SO-FALQON might be relevant to breakdown of adiabaticity \cite{magann2022lyapunov}.
The saturation value of the approximation ratio could be improved by several modifications including addition of random kicks in $\beta_k$, introduction of reference perturbations, and iterative FALQONs as discussed in refs.~\cite{magann2022feedback, magann2022lyapunov}.

\section{conclusions and outlook}\label{conclusion}
In this study we have proposed a new feedback law in FALQON to accelerate convergence to solutions by incorporating the second-order approximation with respect to time interval into approximation of the energy difference in each step.
By numerical simulations on the MAX-CUT problem we have demonstrated that the proposed method scalably reduces the number of layers required by more than an order of magnitude compared to the original version of FALQON.
Our results are expected to be an important step forward in solving combinatorial optimization problems by the feedback-based approach on noisy quantum computers while avoiding classical parameter optimization needed for QAOA.

In the future one could build upon this study to further reduce the circuit depth.
For example, it would be interesting to parametrize the rotation term of the problem Hamiltonian in a similar fashion to QAOA and determine the parameters by a feedback law together with $\beta_k$.
Warm-start approaches would be also an intriguing direction.
In QAOA the number of layers is significantly reduced by biasing the initial state and driver Hamiltonian towards approximate solutions obtained by classical algorithms in advance \cite{egger2021warm, tate2023bridging, okada2024systematic}.
The warm-start technique would be naturally applicable to FALQON and is easily expected to reduce the circuit depth furthermore.

\section*{Acknowledgements}
This work is supported by MEXT Quantum Leap Flagship Program (MEXTQLEAP) Grant Nos. JPMXS0120319794 and JPMXS0118067394, and JST COINEXT Grant No. JPMJPF2014.
K.M. is supported by JST PRESTO Grant No. JPMJPR2019 and JSPS KAKENHI Grant No. 23H03819.

\section*{Appendix: numerical simulation of SO-FALQON without introducing the cap}
\begin{figure}[t]
  \centering
  \includegraphics[scale = 0.45]{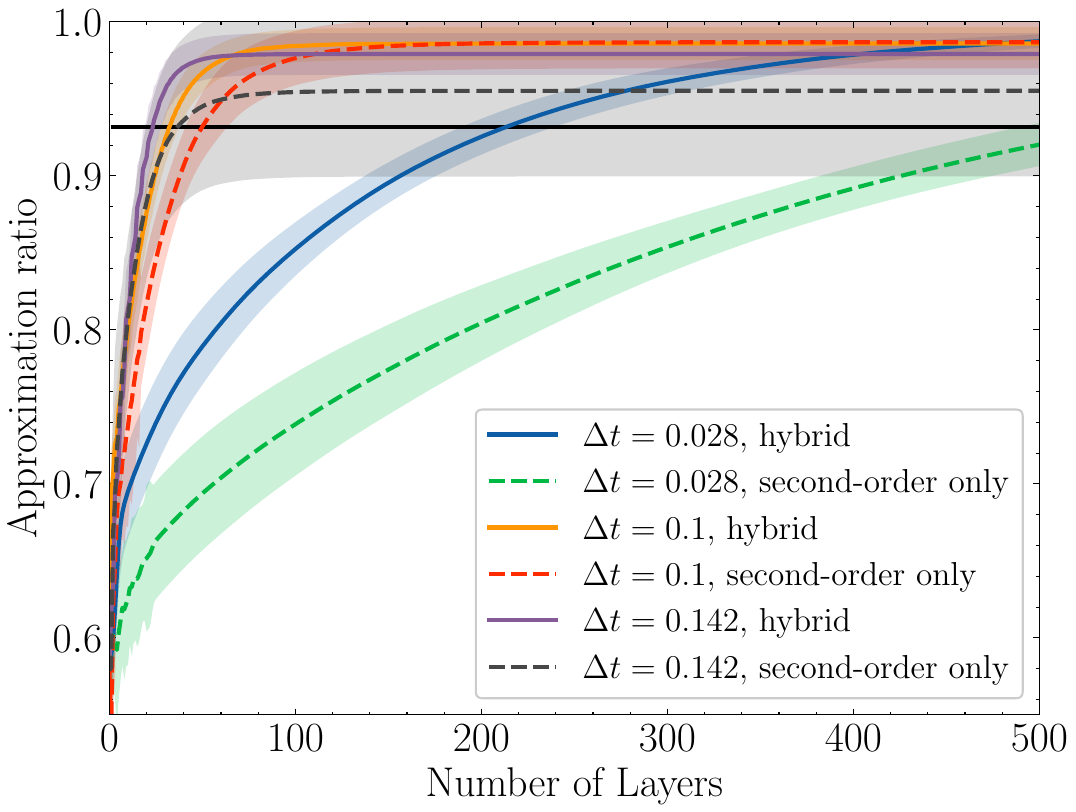}
  \caption{Numerical simulation of SO-FALQON using only the second-order feedback law (dashed line) compared to that making hybrid use of the second-order and first-order feedback laws presented in Sec. \ref{numerics} (solid line). The graph size is $n=12$.}
  \label{pureso}
\end{figure}

Here we present numerical simulation of SO-FALQON using only the second-order feedback law without putting any bound on $\beta_k$. 
Specifically, we adopt Eq. \eqref{newbetak} for $B_k>0$ while changing the sign of the right-hand side of Eq. \eqref{newbetak} for $B_k<0$.
For $B_k=0$, we employ $\beta_k=-(A_{k-1}+C_{k-1}\Delta t)$ to assure a negative value of Eq. \eqref{eq2} in a similar way to the first-order feedback law.
Figure \ref{pureso} shows the simulation results along with those by SO-FALQON making hybrid use of feedback laws presented in Sec. \ref{numerics}.
We can confirm that even in the case of using only the second-order feedback law the approximation ratio almost always monotonically increases and converges to a high value for large $\Delta t$ where FO-FALQON fails to work ($\Delta t=0.1, 0.142$) (see also Fig. \ref{second}).
However, as compared to the hybrid case, the approximation ratio increases more slowly, particularly for $\Delta t =0.028$.

\bibliography{cite}
\end{document}